\documentclass[letterpaper]{article}
\usepackage{aaai}
\usepackage{times}
\usepackage{helvet}
\usepackage{courier}
\usepackage{url}
\usepackage{graphicx}
\usepackage{subfigure}
\usepackage{color}
\usepackage{tablefootnote}
\usepackage{footnote}
\usepackage{balance}
\usepackage{multirow}
\usepackage[authoryear]{natbib}
\usepackage{textcomp}

\newcounter{HWNumberOfComments}
\stepcounter{HWNumberOfComments}

\definecolor{DarkGreen}{rgb}{0.000000,0.6,0.000000}
\newcounter{JSNumberOfComments}
\stepcounter{JSNumberOfComments}

\begin{document}
%
\title{Breaking the News: First Impressions Matter on Online News}

\author{Julio Reis$^{\star}$, Fabr\'icio Benevenuto$^{\star}$, Pedro O.S. Vaz de Melo$^{\star}$, Raquel Prates$^{\star}$, Haewoon Kwak$^{\ddag}$, Jisun An$^{\ddag}$\\~\\
  {\{julio.reis, fabricio, olmo, rprates\}@dcc.ufmg.br}, 
	{\{hkwak, jan\}@qf.org.qa}\\ \\
	$^{\star}$Computer Science Department, Federal University of Minas Gerais, Brazil\\
   $^{\ddag}$Qatar Computing Research Institute, Qatar\\
}

\maketitle
\begin{abstract}
\begin{quote}
A growing number of people are changing the  way they consume  news, replacing the traditional physical newspapers and magazines by their virtual online versions or/and weblogs. The interactivity and immediacy present in online news are changing the way news are being produced and exposed by media corporations. News websites have to create effective strategies to catch people's attention and attract their clicks. 
In this paper we investigate possible strategies used by online news corporations in the design of their news headlines. We analyze the content of 69,907 headlines produced by four major global media corporations during a minimum of eight consecutive months in 2014. In order to discover strategies that could be used to attract clicks, we extracted features from the text of the news headlines related to the sentiment polarity of the headline. We discovered that the sentiment of the headline is strongly related to the popularity of the news and also with the dynamics of the posted comments on that particular news.
\end{quote}
\end{abstract}

\section{Introduction}

A growing number of people are changing the  way they consume  news, replacing the traditional physical newspapers and magazines with their virtual online versions or/and weblogs \citep{Mitchelstein2009,allan2006online,Tewksbury2005}. Online news can be characterized by two key features: \textit{interactivity} and \textit{immediacy} \citep{Karlsson2010}. \textit{Interactivity} relates to the fact that people tend to consume  only the news they are interested in, while  \textit{immediacy} states that people expect to be informed about the latest news with practically no delay.  These characteristics changed the way news are being produced and distributed \citep{paterson2008making}. For instance, the news cycle (i.e. the time lag between when a news organization becomes aware of an issue and publishes it) has been drastically shortened \citep{Karlsson2010}.

Online news industry is competitive. First, there are many more online news sites and second, unlike traditional newspapers, online news sites do not have any physical restriction on the amount of information they can put in, thus they are able to publish more. Given that people are only willing to spend a limited time for consuming news, it is critical for news sites to have effective strategies to catch people's attention and attract their clicks.  However, one aspect that  is not yet fully understood are the strategies adopted by companies in this industry to make the news, and consequently the online newspapers, more attractive to users.

One key aspect related to news dissemination and its attractiveness is the headline that describes it.
A headline's purpose is to quickly and briefly draw attention to the story. The large type front page headline did not come into use until the late $19^{th}$ century when increased competition between newspapers led to the use of attention-getting headlines \citep{evans1974news}. Potentially, headlines determine how many people read the news.
With the increasing competition of the online world, it is not surprising to see headlines become more aggressive, exaggerated, and somewhat misleading.
More important, headlines are not only the first impression of news articles. Recent efforts suggest that they can even drive the way users perceive the rest of the content associated to them by affecting the way people will remember it~\citep{ecker2014effects}.

Despite the absolute importance of headlines on news production and consumption, little is known about them.
A better understanding of the sentiments expressed in headlines, the popularity of news articles as a function of headlines aspects and the kind of comments news items trigger is key for the design of new systems and, ultimately to society.  All this calls for a better knowledge of today's news articles, which we attempt to provide in this paper.
Our effort relies on using sentiment analysis as a mean to capture the strength of sentiments expressed in news headlines (i.e. if they are exaggerated negatively or positively, or if they are neutral). More specifically, we work towards answering the following specific questions:
\begin{itemize}
	\item What is the sentiment of the news produced by popular newspapers? Do they produce more negative, positive or neutral news?
	\item Extremely negative or positive news are able to attract more clicks? In other words, is it worth creating exaggerated headlines?
	\item How are users reactions to news articles with different sentiments associated to them. Does the sentiment polarity of users' comments reflect the polarity of the news headline?
\end{itemize}

To perform our analysis we collected and analyzed 69,907 news produced by four major global media corporations -- The New York Times, BBC, Reuters, and Dailymail -- during a minimum of eight consecutive months in 2014. Then, we designed an experimental methodology to classify each piece of news in terms of its sentiment, its popularity and its topic. Our analysis resulted in a number of interesting findings. We show that the majority of the news produced have negative headlines, and the amount of negative news produced is generally constant over time. The topic category  ``World'' concentrates most of the produced news and is one of the most negative categories. We also show that extremely negative and positive headlines tend to attract more popularity, while neutral headlines tend to be less attractive. Finally, our analysis about users' comments show that comments tend to be negative, independently if the news are positive, negative or neutral. As we will discuss later, these observations have numerous implications to the design of systems that can better support users in choosing and having access to the information they are more interested in. Ultimately, our results contribute to understanding communication in the contemporary media environment, as well as the society we live in.

The rest of the paper is organized as follows. Next, we present a literature review on online news. We then describe the dataset that we collected for the study.
We first study the polarity of news headlines and how it differs across topics and changes over time. We then look at how the sentiment of headline relates to the popularity of the news article. Lastly, we characterize the news comments by their sentiments and how it associate with headlines and news articles. We conclude with a discussion about the implication of our findings.

\section{Literature Review}

Online news has been extensively studied in different domains including media and communication, psychology, and computer science. We briefly describe those efforts regarding news production and consumption with different perspectives.

\textbf{News headlines.} \quad Recently, researchers have conducted an experiment in which factual or opinion news were presented to participants, but with different headlines. The authors conclude that misleading headlines affect readers' memory,  inferential reasoning, and behavioral intentions~\citep{ecker2014effects}. They argue that these effects arise not only because headlines constrain further information processing, biasing readers towards a specific interpretation, but also because readers struggle to update their memory in order to correct initial misconceptions.
These efforts highlight the importance of news headlines. It goes beyond attracting users to read news and it even changes individuals' perceptions or attitudes towards the content, motivating a study dedicated solely on characterizing headlines.

\textbf{Reading Habits.} \quad The ``Mood Management Theory'' \citep{Zillmann} states that individuals' access to media is highly capable of changing their mood states. An experimental study related to reading habits \citep{Biswas} shows that when people have to choose the news they want to read, they are directly influenced by the environment  in which they had been earlier. In this experiment, authors placed men and women in bad and good environments and then provided them with magazine articles. Results showed that those who were exposed to a bad environment were drawn to good news. It also showed that people in a good mood read more bad news than people in a bad mood.
There are recent efforts that attempt to characterize bias on reading habits of online news users and news coverage of specific events.
\cite{an2014sharing} examine aspects related to freedom of reading choices offered by social media sites. However, as cultural habits are deeply ingrained, reading behaviors might be hard to change. Using real instances of political news shared in Twitter they study the predictive power of features related to selective exposure, news source credibility and informativeness, and user socialization. Then, authors  built a prototype of a news sharing application that promotes serendipitous political readings along these dimensions. \cite{esiyok2014users} study reading habits of users of an online news portal to provide insights for better design of news recommendation systems.
As headlines may directly impact reading habits, we hope our study adds a complementary perspective to the above studies.

\textbf{News Coverage and Popularity.} \quad A few studies have used a large scale dataset of online news to understand how features extracted from news articles and their comments are related to news coverage and popularity. \cite{kwak2014understanding} characterize the structure of global news coverage of
disasters and its determinants by using a large-scale news dataset. They find strong regionalism in news geography and identify the type of news that is more likely to be covered globally.
Then, a recent effort proposes a popularity prediction approach to ranking online news~\citep{tatar2014popularity}.
Based on features extracted mainly from users' comments, they formulate a classification problem where the goal is not to infer the precise attention that an article will receive, but to accurately rank articles by their predicted popularity.
We hope our analysis related to news categories, popularity, and sentiments may complement existing studies on news coverage and bring into perspective novel features that are discriminative and that potentially could be incorporated into classifiers.

\textbf{News Propagation.} \quad Propagation aspects were analyzed in \citep{naveed2011bad} and \citep{wu2011does}, using Twitter as a proxy. In the first study, a forecasting model was developed for computing the probability that a given tweet would be forwarded solely based on its content. As made explicit in the very title of the work, they identified evidences that negative news spread faster. In addition, the results provide further interesting insights into propagation dynamics.  The second study focuses on temporal dynamics of information, in order to predict URLs that fade rapidly following their peak of popularity and those that fade more slowly. Using LIWC\footnote{http://www.liwc.net/}  to measure sentiments, they show that the most negative information quickly disappears and the most positive tends to persist. Finally, \cite{hansen2011good} take a qualitative and quantitative approach, also using Twitter, and present evidence that negative sentiment enhances virality in the news segment, but not in the non-news segment.

\textbf{Sentiment analysis of news.} \quad
Recent efforts have explored sentiment analysis to examine news articles or to create novel applications related to this context.
For example, \cite{zhang2010trading} analyze a set of blogs and news articles to propose a method that trades stocks based on sentiment of company-related news.
Another application, namely Magnet News \citep{reis2014magnet}, is a tool that separates positive from negative news in popular newspapers and allows readers to choose which pieces of news to read according to their polarity. More related to our effort, \cite{diakopoulos2011topicality} explore relationship among news comments' topicality, temporality, sentiment, and quality in a dataset of 54,540 news comments. Their observations indicate that comment sentiments, both positive and negative, can be useful indicators of discourse quality. As we will show later, using a different dataset for comments, our findings present a different perspective.

\section{Methodology}

In order to characterize online news headlines, first we collected news from four major news sources. Then we inferred their sentiment, popularity, and topical category. Next we briefly describe the methodology adopted for the analysis.

\subsection{Collecting News from Sources}

We crawled data from four different news sources: BBC News Online\footnote{http://www.bbc.co.uk}, Dailymail Online\footnote{http://www.dailymail.co.uk}, The New York Times\footnote{http://www.nytimes.com} and Reuters Online\footnote{http://www.reuters.com}. All of them are world wide known online news media  with millions of readers daily.
For BBC, Dailymail, and The New York Times we monitored their RSS feed daily from March to November 2014. Articles on the main page of the website were crawled and blogs and news that are featured in specific sections were excluded. 

To crawl Reuters we needed to retroactively collect the newspaper archive, from January 2013 to September 2014. We focused only on the ``Top News'' publications. For this specific source we were also able to crawl the user comments for each piece of news.
Table~\ref{table:datasets} summarizes the main statistics about these four datasets.

\begin{table*}[t] \centering \caption{Data collection statistics per source}\label{table:datasets}
\small
\begin{tabular} {|l|c|c|c|c|c|}
\hline
   & Collection period  & \# News &  \# Comments & Average Popularity   & Mean Popularity\cr \hline
\textsf{BBC News} & Mar/2014-Nov/2014    & 11,707  & - & 671.19 & 67 \cr
\textsf{Dailymail}     & Mar/2014-Nov/2014  & 26,867 &  - & 456.89 & 75 \cr
\textsf{New York Times}     & Mar/2014-Nov/2014  & 4,256 & - & 630.03 & 189 \cr
\textsf{Reuters}     & Jan/2013-Sep/2014  & 27,051 & 166,329 & 149.48 & 7 \cr \hline
\end{tabular} 
\end{table*}

\subsection{Inferring Sentiment}

Recent efforts show that there is no one method that is most suitable for all kinds of data~\citep{PollyannaCOSN}. Thus, instead of using one popular sentiment analysis method, we briefly investigated which method would be the most appropriate for our datasets. To do so we randomly selected 100 news headlines and 100 comments from our data and asked three volunteers from our research lab to manually label these comments as positive, negative or neutral. Each volunteer labeled the data independently, without the influence of the others. The percentage of agreement between the volunteers was 92\% for headlines and 90\% for comments.

We used this small labeled dataset as a ground truth and we compared the performance of eight sentiment analysis methods used by the iFeel system~\citep{AraujoWWW}.
Sentistrength~\citep{sentistrength1} performed better for both, headlines and comments, with an accuracy of 74\% and 79\%, respectively. Other methods showed competitive results with Sentistrengh, but an interesting property of this machine learning based method is that it measures the strength of the polarity sentiments in texts in a scale from -5 (very negative) to +5 (very positive). This provides an interesting perspective of the intensity of the sentiment expressed in the news headlines. Table~\ref{tab:example-title} provides examples of headlines classified according to SentiStrength scores. There was no data classified with scores 1 or -1 in our dataset.

\begin{table*}[t]
  \centering
  \caption{Examples of headlines and their sentiment strength score}
    \begin{tabular}{|c|p{12cm}|} \hline
    Sentiment Score & Headline \\ \hline
    \multicolumn{1}{|c|}{-5} & US Army Fort Hood murder-suicide: Soldier kills three \\ \hline
    \multicolumn{1}{|c|}{-4} & Meat from cattle slaughtered in `cruel' kosher ceremony is in your high street burge \\ \hline
    \multicolumn{1}{|c|}{-3} & The shock troops sent to terrorise Putin's enemies as Crimea prepares to vote on joining Russia \\ \hline
    \multicolumn{1}{|c|}{-2} & Room for a little one? SIX-seater buggy spotted being pushed around Cambridge confusing tourists and locals \\ \hline
    \multicolumn{1}{|c|}{0} & From baronets to bohemians: Britain's heart-throbs of the National Portrait Gallery revealed as blog lists HOTTEST artworks in history \\ \hline
    \multicolumn{1}{|c|}{2} & Ta da! New Guinness World Record set for completing a Rubik's Cube in just 3.253 seconds... by a robot \\ \hline
    \multicolumn{1}{|c|}{3} & J.K Rowling delights Harry Potter fans by posting 2,400 word History Of The Quidditch World Cup on Pottermore \\ \hline
    \multicolumn{1}{|c|}{4} & He's a real-life hero! Liam Neeson saves stray dog from `teenagers throwing stones at it' \\ \hline
    \multicolumn{1}{|c|}{5} & Caravan of love: Britain's youngest parents enjoyed romantic beach break together just a month before their baby was born and gave each other ``True Love'' bracelets \\ \hline
    \end{tabular}
  \label{tab:example-title}
\end{table*}

\subsection{Inferring Popularity}

Ideally, we would like to know how many times each news URL has been clicked on. However, such information is not publicly available, and thus we attempt to approximate the popularity by the number of clicks they received through Bit.ly\footnote{http://bit.ly}. Bit.ly is a well known URL shortening service that shortens millions of URLs daily \citep{Antoniades}. The service API provides the possibility of checking the total number of clicks that a shortened link has received. This strategy provides an interesting estimate of popularity as the service keeps the same shortened URL for requests of the same long URL. For example, if a user requests to shorten \url{www.dummy.com}, Bit.ly will return a shortened URL such as \url{http://bit.ly/dummyID}. But, if another user tries to shorten the same URL, the system will also return \url{http://bit.ly/dummyID} \footnote{Users may customize their shortened URL, however, if they do, the system will keep the association to the original long URL in the server and the API will return the total number of clicks on all shortened URLs (generated by the system or customized) associated to the long URL.}. The number of clicks was collected 20 days after we finished collecting the news. Considering that bit.ly URLs normally have 60\% of their clicks on the first day after being released \citep{Antoniades} and that  recent evidence \citep{castillo@cscw2014} suggests that five days are enough to approximate the final number of visits to news articles of Al Jazeera, we argue that a threshold of 20 days provides us with a safe approximation of the final number of clicks. Table~\ref{table:datasets} summarizes the overall popularity numbers for the news collected per source.

\subsection{Categorizing News}

In order to infer the topical categories of the news articles, we use meta information embedded in the news URLs. 
For example, the URL \url{http://www.nytimes.com/2014/03/29/sports/ncaabasketball/once-again-izzo-is-one-step-away.html}  is  categorized  as ``Sports''. 
BBC, Dailymail and The New York Times adopt such a strategy in their URLs, and thus it was possible for us to parse their URL to infer the topic of their news articles.
For Reuters, the news was grouped by category in the archive from which we collected the data. So we took the topic from the category name.
In total, only  9\% of the news items from this online newspaper could not be identified and was discarded for this analysis. Table~\ref{table:category} shows the distribution of the number of news items for each newspaper across the different categories we identified. We have standardized the names of some categories to simplify the comparison among news sources. We note that World news is the most representative category in all news sources.

\begin{table*}[htbp]
  \centering
  \caption{Distribution of news gathered across category}
    \begin{tabular}{|r|r|r|r|r|r|r|r|}
		\hline
    \multicolumn{2}{|c|}{\textsf{BBC News}} & \multicolumn{2}{|c|}{\textsf{Dailymail}} & \multicolumn{2}{|c|}{\textsf{New York Times}} & \multicolumn{2}{|c|}{\textsf{Reuters}} \\ \hline
    \multicolumn{1}{|l}{Category} & \multicolumn{1}{r|}{\%} & \multicolumn{1}{|l}{Category} & \multicolumn{1}{r|}{\%} & \multicolumn{1}{|l}{Category} & \multicolumn{1}{r|}{\%} & \multicolumn{1}{|l}{Category} & \multicolumn{1}{r|}{\%} \\ \hline
    \multicolumn{1}{|l}{World } & \multicolumn{1}{r|}{26.8} & \multicolumn{1}{l}{World} & \multicolumn{1}{r|}{69.2} & \multicolumn{1}{l}{World} & \multicolumn{1}{r|}{36.4} & \multicolumn{1}{l}{World} & \multicolumn{1}{r|}{64.3} \\
    \multicolumn{1}{|l}{Sport} & \multicolumn{1}{r|}{10.4} & \multicolumn{1}{l}{Science \& Tech} & \multicolumn{1}{r|}{7.4} & \multicolumn{1}{l}{Business} & \multicolumn{1}{r|}{8.5} & \multicolumn{1}{l}{Politics} & \multicolumn{1}{r|}{4.0} \\
    \multicolumn{1}{|l}{Business} & \multicolumn{1}{r|}{5.8} & \multicolumn{1}{l}{Femail} & \multicolumn{1}{r|}{7.4} & \multicolumn{1}{l}{Sports} & \multicolumn{1}{r|}{7.9} & \multicolumn{1}{l}{Tech} & \multicolumn{1}{r|}{3.0} \\
    \multicolumn{1}{|l}{Entertainment} & \multicolumn{1}{r|}{4.5} & \multicolumn{1}{l}{Health} & \multicolumn{1}{r|}{6.1} & \multicolumn{1}{l}{Upshot} & \multicolumn{1}{r|}{2.4} & \multicolumn{1}{l}{Deals} & \multicolumn{1}{r|}{1.3} \\
    \multicolumn{1}{|l}{Health} & \multicolumn{1}{r|}{2.9} & \multicolumn{1}{l}{Travel} & \multicolumn{1}{r|}{5.5} & \multicolumn{1}{l}{Technology} & \multicolumn{1}{r|}{2.1} & \multicolumn{1}{l}{Aerospace \& Defense} & \multicolumn{1}{r|}{0.7} \\
    \multicolumn{1}{|l}{Technology} & \multicolumn{1}{r|}{2.7} & \multicolumn{1}{l}{Money} & \multicolumn{1}{r|}{1.5} & \multicolumn{1}{l}{Arts} & \multicolumn{1}{r|}{1.6} & \multicolumn{1}{l}{Money} & \multicolumn{1}{r|}{0.7} \\
    \multicolumn{1}{|l}{Science} & \multicolumn{1}{r|}{2.1} & \multicolumn{1}{l}{Debate} & \multicolumn{1}{r|}{1.2} & \multicolumn{1}{l}{Science} & \multicolumn{1}{r|}{1.3} & \multicolumn{1}{l}{Health} & \multicolumn{1}{r|}{0.4} \\
    \multicolumn{1}{|l}{Education} & \multicolumn{1}{r|}{1.7} & \multicolumn{1}{l}{Sport} & \multicolumn{1}{r|}{0.3} & \multicolumn{1}{l}{Health} & \multicolumn{1}{r|}{0.8} & \multicolumn{1}{l}{Sports} & \multicolumn{1}{r|}{0.4} \\
    \multicolumn{1}{|l}{Others} & \multicolumn{1}{r|}{41.2} & \multicolumn{1}{l}{Others} & \multicolumn{1}{r|}{1.4} & \multicolumn{1}{l}{Others} & \multicolumn{1}{r|}{38.8} & \multicolumn{1}{l}{Others} & \multicolumn{1}{r|}{25.3} \\ \hline
    \end{tabular}%

  \label{table:category}%
  \vspace{-4mm}
\end{table*}%

\vspace{-2mm}

\subsection{Potential Limitations}

A possible bias of the data collected is that popularity is being inferred based on the number of clicks on bit.ly. We cannot guarantee what led to the click was the news articles' headlines. Users could share URLs in other forums, such as social networks or emails, and present the news differently from the headline, going as far as changing not only the words, but also the sentiment associated to the original headline.

Therefore, to verify the extent to which this happens, we gathered a small sample of data from Twitter, the most important social system in which users shorten their URLs using Bit.ly \citep{Antoniades}.
We randomly selected 100 crawled URLs and, through the official Twitter API
\footnote{https://dev.twitter.com}
, we collected tweets that shared these URLs (shortened). In total, we collected 5,182 tweets.
We noted that 49\% of these tweets contained as text the exact headline of the news item of the crawled URL, meaning that the news headline and tweet corpus are not identical in 51\% of the cases.

Then, we contrasted the sentiment polarity of these tweets with the sentiment polarity of the news headlines to verify the consistency between them (e.g. both are positive, neutral or negative).
In order to explain the results obtained we will make use of a
confusion matrix~\citep{provost1998guest}, depicted in Table~\ref{tab:tweet}.
Each position in this matrix represents the percentage of news headlines in each polarity class (positive, negative and neutral), and how they were classified according to the text they were shared in Twitter. For all classes, a percentage of agreement (i.e. values in the diagonal) is greater than 72\%, indicating a high consistency between the polarities of the headlines and text describing the news articles in Twitter.

\vspace*{-2mm}

\begin{table}[h]
\centering
\small
    \begin{tabular}{|c|c|ccc|} \hline
      \multicolumn{2}{|c}{} & \multicolumn{3}{|c|}{\textit{Tweet}} \\
     \multicolumn{2}{|c|}{}   & Negative & Neutral & Positive \\ \hline
    \textit{~~ ~~} & Negative & \textbf{73.80\%}       & 19.23\%   & 7.69\%        \\
		\textit{Headline} & Neutral & 19.44\%         & \textbf{72.22\%}    & 8.33\%     \\
    \textit{~~ ~~} & Positive & 4.55\%       & 13.64\%    & \textbf{81.82\%}      \\ \hline
    \end{tabular}
		\caption{Confusion matrix for headlines and tweets}
    \label{tab:tweet}
\end{table}

\vspace*{-2mm}

Finally, inferring popularity from Bit.ly can have other forms of bias, for example, towards the news that are more likely to quickly spread through social networks. In order to allow our experiments to be reproduced and the extent of this and any other limitations be quantified by researchers with access to other methods to infer news popularity we have made our dataset available \footnote{http://www.dcc.ufmg.br/\texttildelow{}fabricio/}. 

\section{Headlines Polarity}

\begin{figure}[!h]
	\centering
		\includegraphics[width=0.45\textwidth]{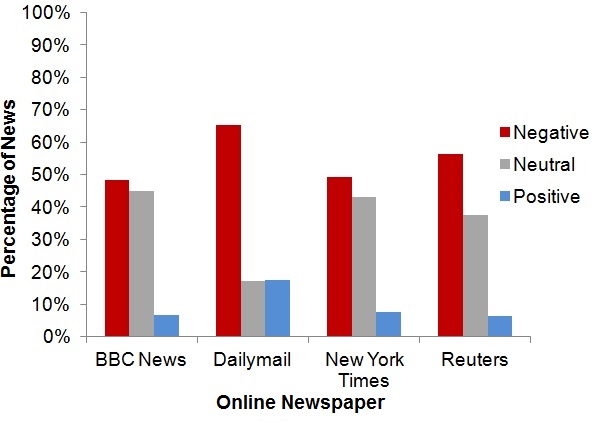}
	\caption{Sentiment polarity of news headlines}
	\label{fig:numberNewsScore}
\end{figure}

We began by measuring the polarity of all news headlines we collected. Figure~\ref{fig:numberNewsScore} shows the amount of news headlines classified as negative, neutral, and positive. We can observe a trend in all four news sources: the negative news headlines are the majority, followed by neutral and positive ones. Particularly, for Dailymail, the amount of negative news headlines (65\%) is extremely higher than neutral and positive ones (the rest 35\%). In order to better understand this result, for all news sources analyzed we decided to break down our analysis into two perspectives: topical, to check if specific categories are more negative than others,
and temporal, to verify if particular seasons or events affect the news headlines sentiment scores.

\begin{figure}[t]
  \centering {
    \subfigure[BBC News]{\includegraphics[width=0.23\textwidth]{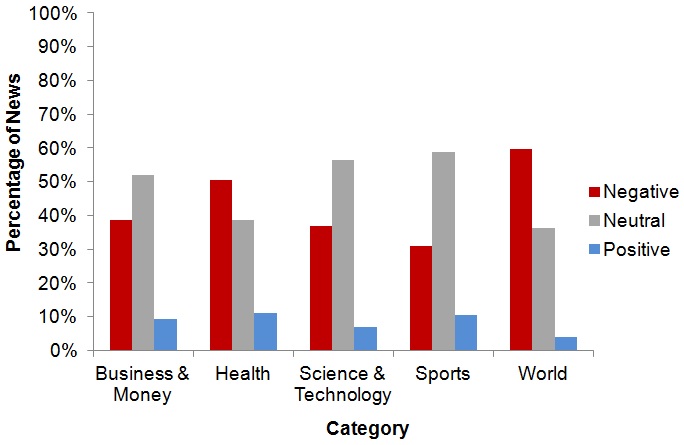}\label{fig:category_bbc}}
		\subfigure[Dailymail]{\includegraphics[width=0.23\textwidth]{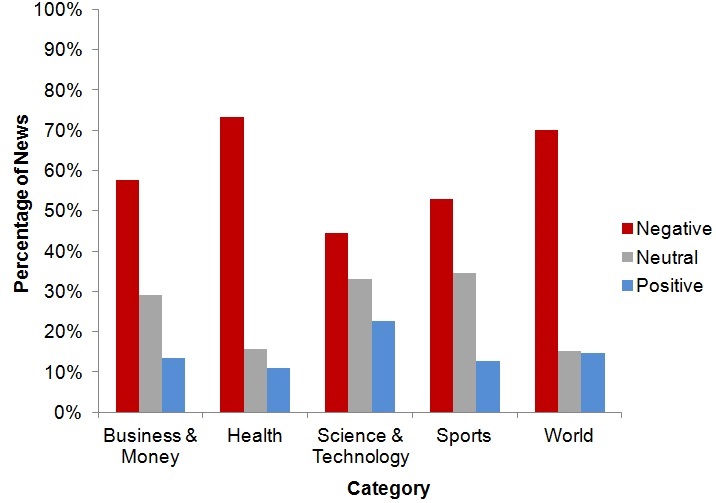}\label{fig:category_dailymail}}
		\subfigure[New York Times]{\includegraphics[width=0.23\textwidth]{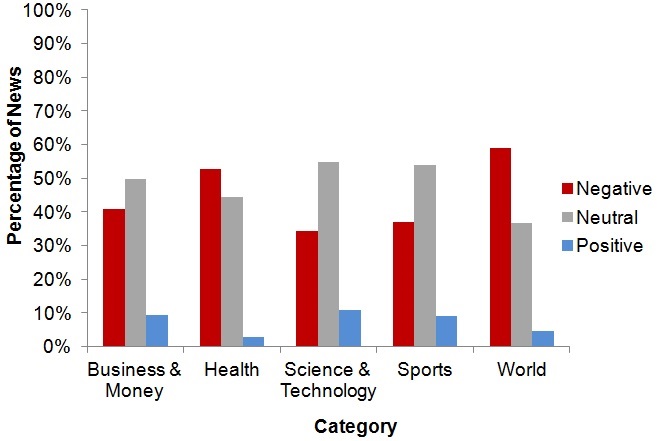}\label{fig:category_nytimes}}
		\subfigure[Reuters]{\includegraphics[width=0.23\textwidth]{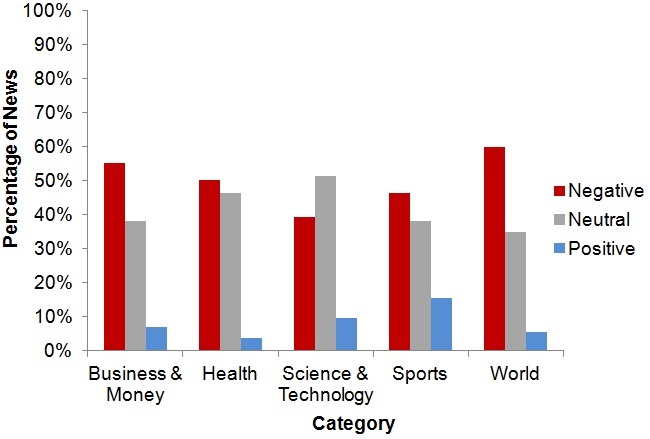}\label{fig:category_reuters}}
		
}
  \caption{Sentiment polarity of news headlines per category}
	\label{fig:categoryNews}
\end{figure}

\begin{figure*}[th!]
  \centering {
    \subfigure[BBC]{\includegraphics[width=0.23\textwidth]{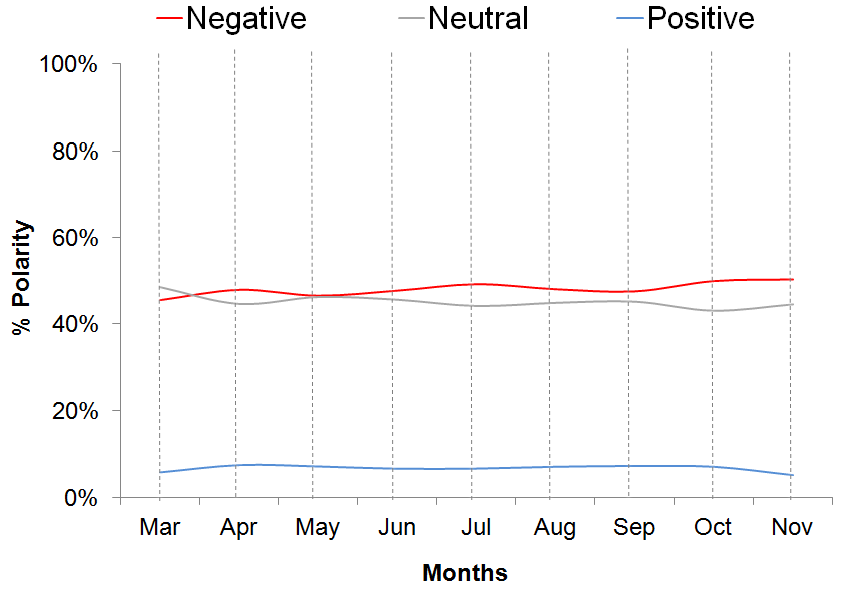}\label{fig:month_bbc}}
		\subfigure[Dailymail]{\includegraphics[width=0.23\textwidth]{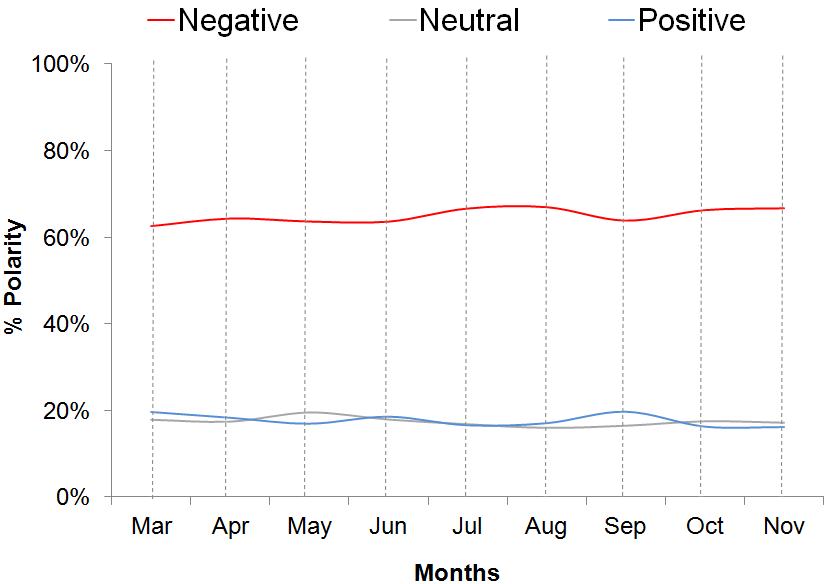}\label{fig:month_dailymail}}
		\subfigure[New York Times]{\includegraphics[width=0.23\textwidth]{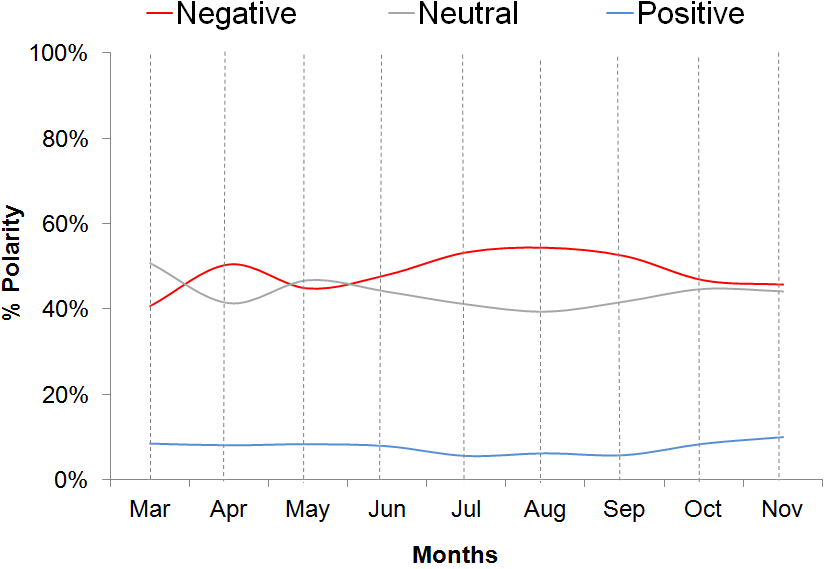}\label{fig:month_nytimes}}
		\subfigure[Reuters]{\includegraphics[width=0.29\textwidth]{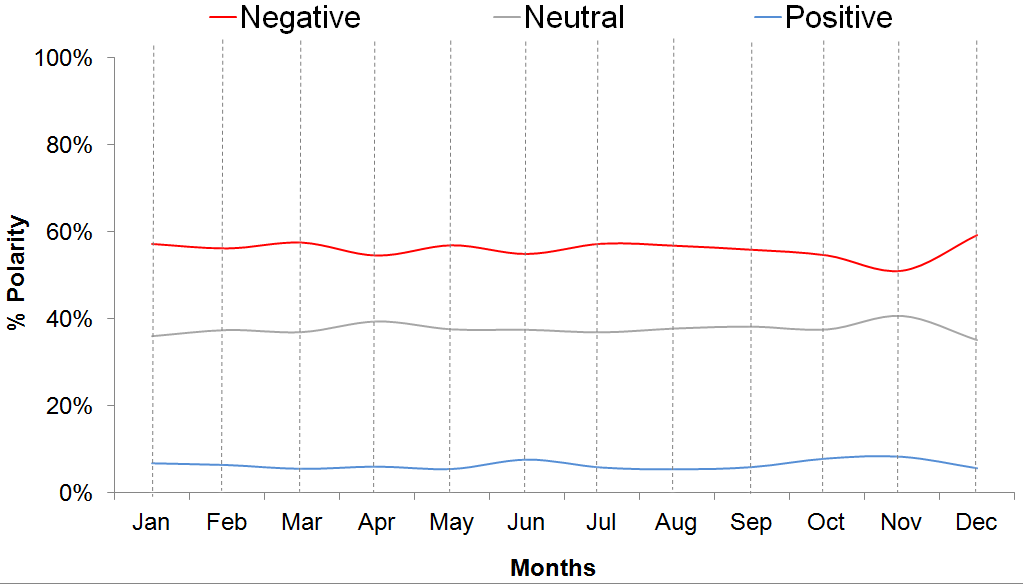}\label{fig:month_reuters}}
}
  \caption{Sentiments over months}
	\label{fig:month_sentiment}
\end{figure*}

\begin{figure*}[th!]
  \centering {
    \subfigure[BBC]{\includegraphics[width=0.24\textwidth]{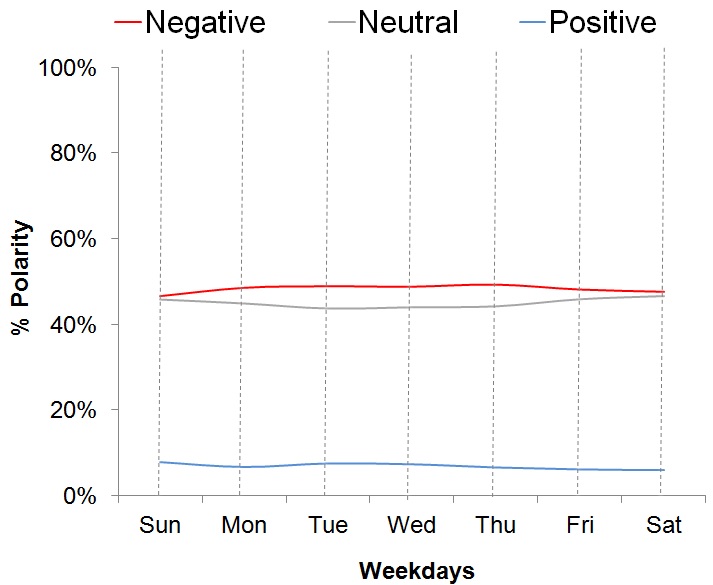}\label{fig:weekdays_bbc}}
		\subfigure[Dailymail]{\includegraphics[width=0.24\textwidth]{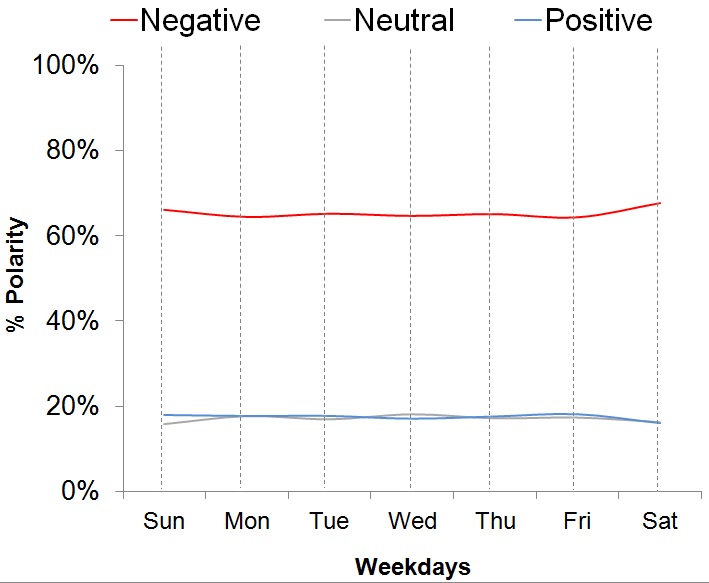}\label{fig:weekdays_dailymail}}
		\subfigure[New York Times]{\includegraphics[width=0.24\textwidth]{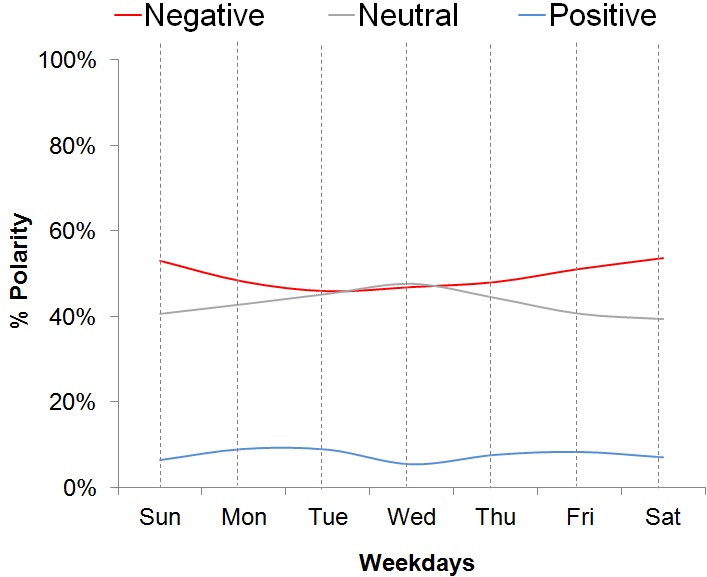}\label{fig:weekdays_nytimes}}
		\subfigure[Reuters]{\includegraphics[width=0.24\textwidth]{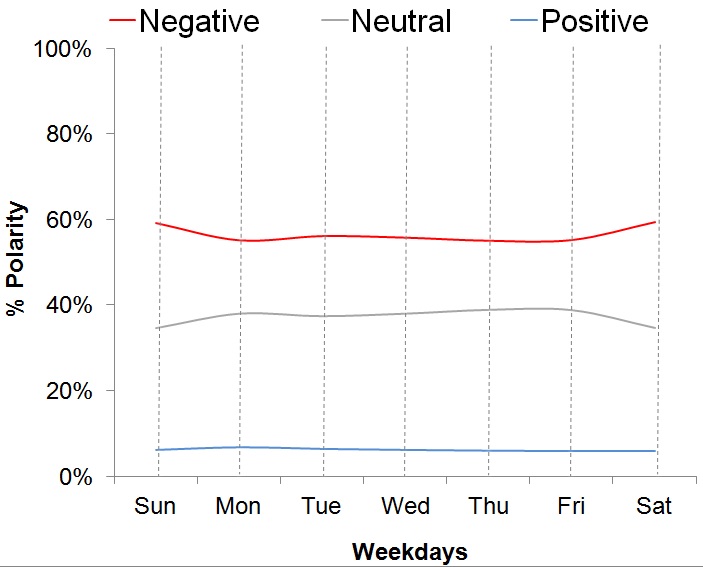}\label{fig:weekdays_reuters}}
}
  \caption{Sentiment per day of the week}
	\label{fig:weekdays_sentiment}
\end{figure*}

\subsection{Sentiment across Categories}

In this section, we look into the sentiment scores of news headlines across different categories. We analyzed the sentiment for all news headlines and compared the categories that were present in all four online news\footnote{Health and Sports are present in all four sources. To generate the other three categories we grouped similar categories in different sources. The categories ``Business and Money" and ``Science and Technology" grouped categories containing these terms in the sources. It is worth noting that World was present in all sources except Dailymail. A manual verification of the contents of Dailymail news was done and the category denominated News was considered to be equivalente to what the other sources denominated World.}:  Business and Money, Health, Science and Technology, Sports and World. The results are depicted in Figure~\ref{fig:categoryNews}. First, observe that different categories have different distributions of sentiment scores. Health and World are predominantly negative for all four news sources. Considering that World is the most popular category in the news sources we analyzed (Table~\ref{table:category}), it may be one of the reasons for the overall negative trend observed in Figure~\ref{fig:numberNewsScore}. Except for Dailymail, the Science and Technology category tends to have more neutral news headlines, but still the amount of negative ones is higher than the positive ones. Looking at all the categories, only a few unpopular ones (that are grouped in the category Others in Table~\ref{table:category}) have more positive than negative news headlines.
For instance, the categories ``Dining'' and ``Travel'' of The New York Times have essentially positive headlines as they are dedicated to review and suggest places to go.
Next, we investigate if the overall trend toward negative news headlines is consistent over time. 

\subsection{Sentiment Score along Time}

We analyzed headlines sentiment scores as a function of time in relation to three aspects. First, we looked if news headlines sentiment changed along the months for the data collected. Next, if different days of the week or events/hot topics influenced sentiment polarity of headlines. Looking at the period analyzed, we can observe from Figure~\ref{fig:month_sentiment} that  the number of negative news headlines is constantly higher, with slight fluctuations specially in March for BBC and The New York Times. Also, the analysis per day of the week did not show much change in the distribution of the sentiment scores along the week, except for a slight increase in the number of negative headlines on weekends for the four sources (see Figure \ref{fig:weekdays_sentiment}).

Moreover, by looking at Figure~\ref{fig:month_sentiment}, we do not notice any spikes in any of the curves. Therefore, it is possible to conclude that events that have taken place during this period have not influenced the sentiment of headlines in general, that is, when we aggregate all categories. However, if we look at each category separately, we noticed spikes for a short period of time, which means that they may be influenced by events or hot topics.

\begin{figure}[t]
  \centering {
    \subfigure[Dailymail]{\includegraphics[width=0.23\textwidth]{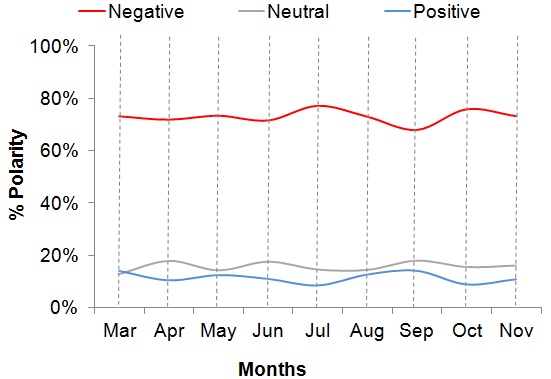}\label{fig:health_dailymail}}
		\subfigure[New York Times]{\includegraphics[width=0.23\textwidth]{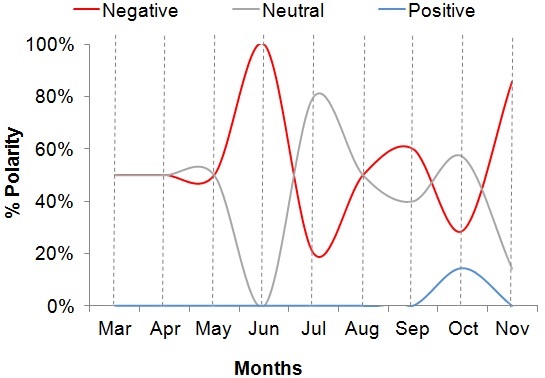}\label{fig:health_nytimes}}
}
  \caption{Sentiment of Month - Health}
	\label{fig:month_nytimesdailymail}
\end{figure}

In all categories spikes can be seen for at least one of the sources. However, the spikes distribution in any category from a source tend to be very different from the other sources. It is also interesting to note that, for all categories and sources, the positive sentiment headline does never, in any period of time, achieve the highest percentage. The highest percentage for headline sentiment at a given moment is either negative or neutral. The World category is the only one in which the negative headlines are totally predominant  (the percentage of the neutral headlines is only higher for a short period in the beginning of March for The New York Times). Due to a space limitation, we only show the headline sentiment distribution for the Health category for the Dailymail and New York Times in Figure~\ref{fig:month_nytimesdailymail}.

\section{Sentiment Score versus Popularity}

\begin{figure}[t]
  \centering {
    \subfigure[BBC News]{\includegraphics[width=0.23\textwidth]{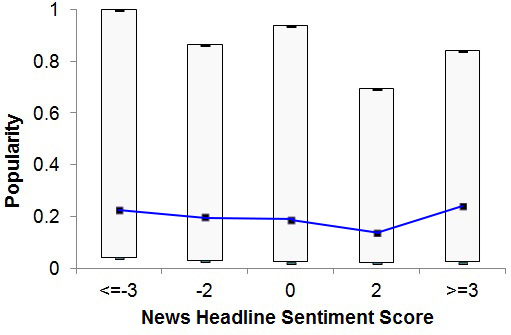}\label{fig:popularity_bbc}}
		\subfigure[Dailymail]{\includegraphics[width=0.23\textwidth]{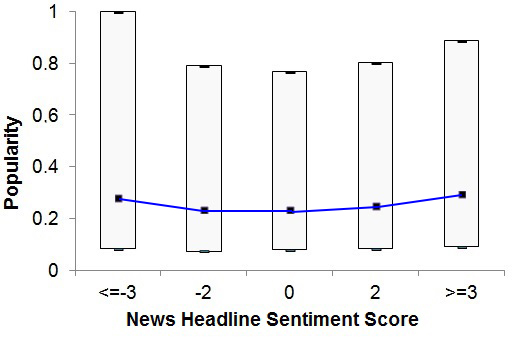}\label{fig:popularity_dailymail}}
		\subfigure[New York Times]{\includegraphics[width=0.23\textwidth]{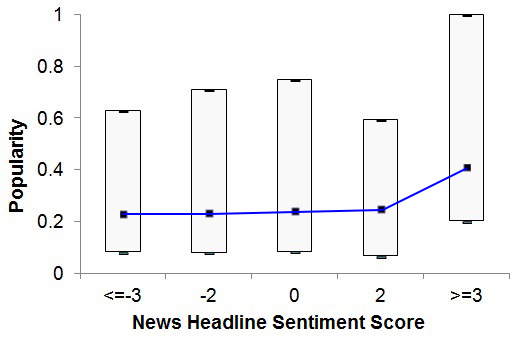}\label{fig:popularity_nytimes}}
		\subfigure[Reuters]{\includegraphics[width=0.23\textwidth]{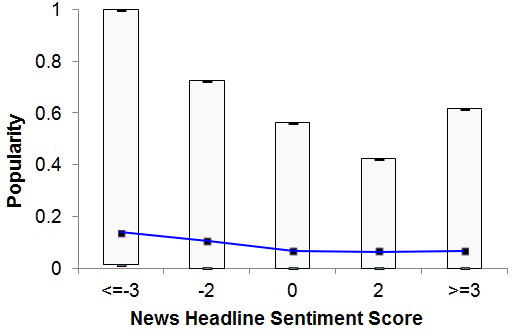}\label{fig:popularity_reuters}}
}
  \caption{Popularity of headlines by sentiment score}
	\label{fig:popularityMediana-NumberNews}
\end{figure}

Next, we investigated the relationship between popularity and the headlines' sentiment score.
Figure~\ref{fig:popularityMediana-NumberNews} shows the news popularity as a function of the sentiment score of the headlines for all four news sources.
More specifically, it shows the average values of popularity for news headlines that have a given sentiment score. We normalized the popularity numbers to allow a better comparison among the four news sources and we  grouped the extreme values of sentiment strength from -5 to -3 and from 3 to 5 in order to have at least 100 data points in taking the mean. The box plot consists of the mean values and the first and third quartiles. Observe that for all news sources, an extreme sentiment score obtained the largest mean popularity. Particularly, for both BBC and Dailymail, the extreme negative and the extreme positive values were associated with the most popular news articles. This suggests that strongly negative or strongly positive news tend to be more attractive to Internet users.

This finding is in line with observations made for meme propagation in social networks \citep{coscia2014average}. The authors investigated what the meme's success factors are. Based on text similarity, they realized that successful memes are located in the periphery of the meme similarity space. Interestingly, our results suggest that an headline has more chance to be successful if the sentiment expressed in its text is extreme, towards the positive or the negative side. Results suggest that neutral headlines are usually less attractive.

\section{Characterizing the Sentiment of Comments}

Finally, we characterize comments posted to Reuters\footnote{Reuters was the only source that had comments available (see Table~\ref{table:datasets}).} articles aiming at answering the following questions.
What is the expected percentage of hostile comments on news? Are there categories of news that attract more negative comments? Do headlines with extreme sentiment scores attract more comments? What is the expected sentiment of these comments? 

\begin{figure}[t]
	\centering
		\includegraphics[width=0.45\textwidth]{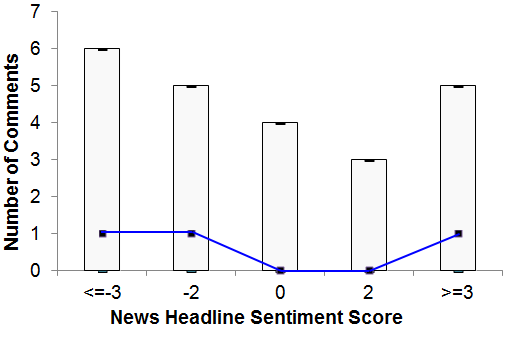}
	\caption{Boxplot for the number of comments the news headlines of a given sentiment score received}
	\label{fig:commentsVsTitle}
\end{figure}

First, in Figure~\ref{fig:commentsVsTitle}, we show the boxplot for the number of comments the news headlines of a given sentiment score received in our dataset. Again, the bottom of the bar corresponds to the first quartile (0 for all sentiment scores), the square inside the bar represents the median, and the top of the bar corresponds to the third quartile. Observe that news with negative headlines are more likely to receive comments than others. Moreover, note that news tend to receive more comments as the sentiment score of the headline moves away from neutral (0). This corroborates the results we showed in the previous section, indicating that news with neutral headlines are also less attractive to users as they tend to trigger less interactions among users.

\begin{figure}[t]
	\centering
		\includegraphics[width=0.45\textwidth]{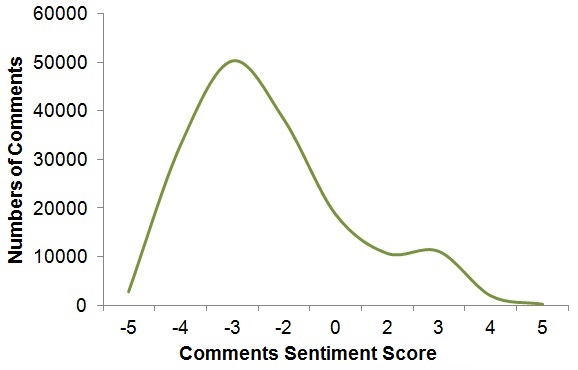}
	\caption{Number of the comments of a given sentiment score}
	\label{fig:scoreComments}
\end{figure}

Given that news with negative headlines tend to receive more comments, we ask: what is the expected sentiment score of the comments on the news? To answer this question, we calculated the sentiment score of every comment in our dataset and we plotted the histogram of comments of a given sentiment score, depicted in Figure~\ref{fig:scoreComments}. Observe that, for this particular online newspaper, the vast majority of comments have negative sentiment scores. In summary, $75\%$ of the comments have a negative sentiment score, $11\%$ are neutral, and $14\%$ have a positive sentiment score. This corroborates the common knowledge and the articles by Krystal D'Costa~\citep{DCosta2013} and Tauriq Moosa~\citep{Moosa2014}, which state that the comments section is usually home for hostility. However, they disagree from results reported in \cite{diakopoulos2011topicality} which indicate that user behavior may vary according to the environment.

In order to verify if the sentiment score of the comment is correlated to (or is a consequence of) the sentiment score of the headline, Table~\ref{tab:matrix} presents the confusion matrix between the sentiment score of the headline and the sentiment score of the comment posted on that headline. We find that negative comments are posted independently of the sentiment score of the headlines. The proportion of negative comments is significantly higher than the others for all sentiment scores of headlines. Even when the headline is summarizing an extremely positive story, the comments are expected to be negative, more than $70\%$ of the times. Neutral headlines are the ones which receive the highest percentage of negative comments. Finally, it is important to point out that in our dataset more than $85\%$ of the users have posted only one comment per news on average, that discards the possibility of this result being biased towards users who comment a lot in a particular news article.

\begin{table}[h]
\centering
\small
    \begin{tabular}{|c|c|ccc|} \hline
      \multicolumn{2}{|c}{} & \multicolumn{3}{|c|}{\textit{Comments}} \\
     \multicolumn{2}{|c|}{}   & Negative & Neutral & Positive \\ \hline
    \textit{~~ ~~} & Negative & \textbf{71.30\%}       & 12.71\%   & 16.00\%        \\
		\textit{News Headlines} & Neutral & \textbf{76.55\%}         & 10.46\%    & 12.99\%     \\
    \textit{~~ ~~} & Positive & \textbf{70.43\%}       & 11.37\%    & 18.19\%      \\ \hline
    \end{tabular}
		\caption{Confusion matrix between the sentiment score of the headline and the sentiment score of the comment posted on that headline}
    \label{tab:matrix}
\end{table}

\begin{figure}[t]
	\centering
		\includegraphics[width=0.5\textwidth]{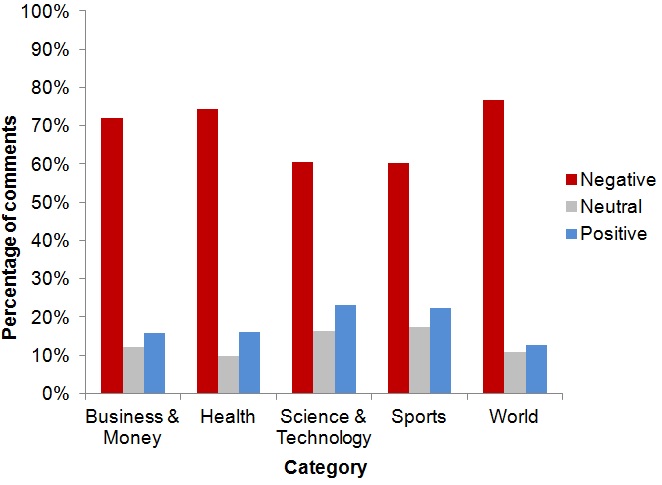}
	\caption{Sentiment of comments of different categories}
	\label{fig:comment_category}
\end{figure}

Lastly, we address the last question: are there news categories that attract negative comments more than others? Figure~\ref{fig:comment_category} shows the amount of comments classified as negative, neutral, and positive across categories. Negative comments are posted significantly more than neutral or positive comments in all categories. Note that \textit{World} and \textit{Health} news are the ones which attract negative comments the most, while \textit{Sports} and \textit{Science and Technology} news are the least ones receiving negative comments. Nevertheless, even \textit{Science and Technology} news receives much more negative comments ($\approx 60\%$) than others. We offer possible explanations of why people leave negative comments in the discussion section.

\section{Discussion}

In the previous section we presented the results of our analysis. We now discuss implications of our findings and also suggest directions for future research.

\subsection{Effects of News Headlines on User Perception and Behavior}

Psychologists have long known that first impressions matter~\citep{digirolamo1997first}.
Headlines give the first impression of news articles and they can drive how users perceive their content. By drawing attention to certain details or facts, a headline can affect which existing knowledge is activated in one's brain. By its choice of phrasing, a headline can influence one's mindset so that readers later recall details that coincide with what they were expecting, leading individuals to perceive the same content differently according to the headline~\citep{dooling1971effects}.

In order to briefly look at the relation between the sentiment expressed in headlines and content,
Table~\ref{tab:textVsHeadlines} presents a confusion matrix for the two variables. We find that there is a non-negligible amount of news in which the sentiment expressed in headlines is different from the one expressed in the content. This comparison is simplistic since content is longer and tends to contain both positive and negative sentences (even when it is negative), whereas the headline is concise and usually expresses only one aspect of the content. It is also curious that positive content is more often represented by neutral or negative headlines than positive ones, whereas negative and neutral content are more often represented with headlines of the same polarity.  Thus, an interesting direction for future research would be to conduct a user study in order to investigate whether users perceive this gap between headlines and content sentiment polarity and what is their attitude towards it.

\begin{table}[h]
\centering
\small
    \begin{tabular}{|c|c|ccc|} \hline
      \multicolumn{2}{|c}{} & \multicolumn{3}{|c|}{\textit{News Headlines}} \\
     \multicolumn{2}{|c|}{}   & Negative & Neutral & Positive \\ \hline
    \textit{~~ ~~} & Negative & \textbf{61.15\%}       & 29.72\%   & 9.13\%        \\
		\textit{News Content} & Neutral & 32.90\%         & \textbf{62.37\%}    & 4.73\%     \\
    \textit{~~ ~~} & Positive & 31.70\%       & \textbf{41.38\%}    & 26.92\%      \\ \hline
    \end{tabular}
		\caption{Confusion matrix for News Headlines and News Content}
    \label{tab:textVsHeadlines}
\end{table}

Another implication from our analysis is related to the predominance of negative news headlines in all the news sources we analyzed. It is known that the news people read have impacts on their behavior \citep{Nguyen} and ultimately can affect society. However, as we noted from the analysis of the news categories, the category  ``World'' contains most of the content generated by the news sources analyzed and is one of the most negative categories. This result seems to support the conjectures that news media usually report foreign disasters and keep a negative tone when describing happenings in other countries as an attempt to emphasize the safety of the home country \citep{leetaru2011culturomics}. Other studies propose to investigate and quantify a possible ideological bias by news organizations \citep{budak2014fair}.

\subsection{Implications for System Design}

There are multiple strategies that could be designed to attract clicks to news items that rely mostly on dissemination mechanisms \citep{naveed2011bad,wu2011does} and recommendation of personalized content \citep{esiyok2014users}.  By characterizing the sentiment expressed in headlines we show that neutral news are usually less popular than extremely positive and negative news. This observation can be leveraged to construct a number of support systems related to online news.

\begin{figure}[h]
	\centering
		\includegraphics[width=0.5\textwidth]{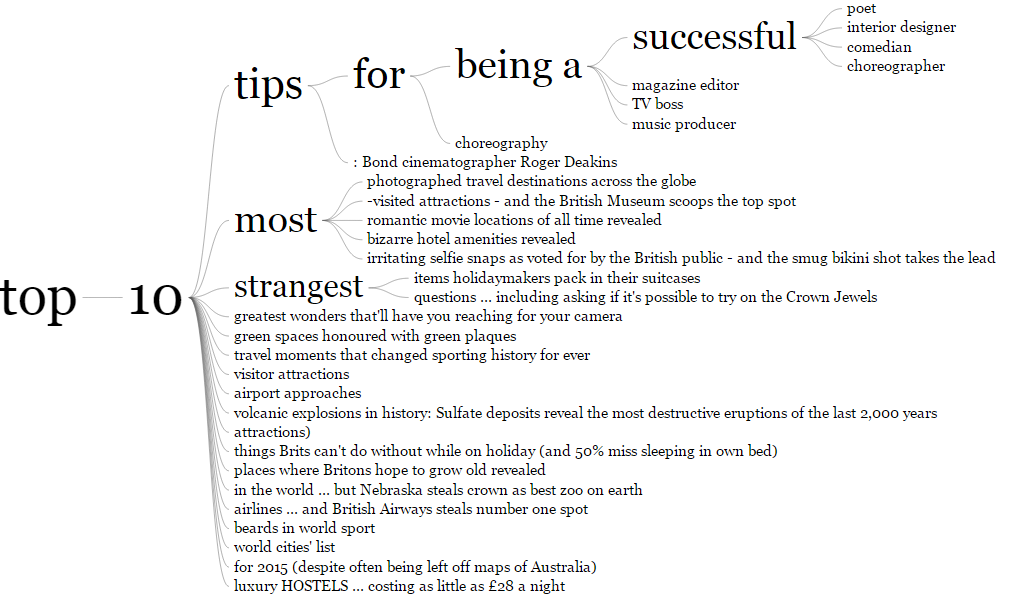}
	\caption{Word tree for root term: Top 10}
	\label{fig:word_tree}
\end{figure}

As an example, a recent work proposes a ranking scheme based on the news popularity prediction \citep{tatar2014popularity}. Our findings related to news popularity show that sentiment scores inferred from headlines are valuable features for popularity prediction. Intuitively, we believe there might be other patterns and features to be extracted from headlines that can help on predictive tasks. For example, we noted a number of headlines that enumerates a sequence of things such as \emph{Top N...} or \emph{Eleven things you should know about}, etc, and that they usually seem to be more popular than the average. Figure~\ref{fig:word_tree} shows a Word tree visualization~\citep{wattenberg-2008-word} for the root term \emph{Top 10} using as input the headlines of our dataset. The visualization shows phrases that branch off from the root term across all headlines in our dataset. A larger font size means that the word occurs more often. One direction we plan to take as future work is to identify patterns on headlines that can help in predicting their future popularity.

\subsection{Why are Comments So Negative?}

How many times have you read or heard the following sentence: ``Don't read the comments.''? Recently, the anthropologist Krystal D'Costa put an exclamation mark to this phrase on an article she wrote for Scientific American: ``Don't read the comments! (Why do we read the online comments when we know they'll be bad?)''~\citep{DCosta2013}. She explains that when people from different backgrounds and cultures interact, they may read something that challenges their world view and then logic and reason disappear: all that matters is that they let you know how wrong you are or the topic/perspective/subject is. All of a sudden, the comments section has become a rant. Tauriq Moosa went further by writing in an article for The Guardian that the comments section ``groans as hatred expands its force, waiting for any point of dissent to break it´´ to unleash its full fury on targets who dare convey some measure of civility or dissent''~\citep{Moosa2014}.

Our findings show that negative comments are ubiquitous. Online news are likely to attract negative comments independently of the sentiment of the headline or the category of the news.
This seems to be related to the situation identified by \cite{cheng2014community}, in which negative comments tend to ``snowball". They claim that negative comments can be seen as a negative feedback and ``authors of negatively-evaluated content contribute more, but also their future posts are of
lower quality, and are perceived by the community as such. Then, these authors are more likely to subsequently
evaluate their fellow users negatively, percolating these effects through the community. In contrast,
positive feedback does not carry similar effects, and neither encourages rewarded authors to write more, nor improves
the quality of their posts.'' This could be one of the reasons why trends of negative comments are always dominating trends of positive ones. It is worth mentioning that Reuters was the only news source from the four sources we considered that had comments available for collection at the time we gathered our dataset (end of November). Today, Reuters has shut down its comments feature and does not allow users to comment on articles anymore. We could speculate that the decision might be related to the predominance of negative comments.

\section{Conclusion}

In this paper we presented a thorough characterization of online news articles based on a dataset of 69,907 news produced by four major global media corporations, gathered during a minimum of eight consecutive months in 2014. We presented the methodology we designed to conduct our analysis putting headlines as a central artifact  as they are key to attract attention in online news. Our methodology used sentiment analysis as a mean to understand intrinsic aspects of headlines and also a strategy to infer news popularity based on a URL shortening service. Our results unveil a number of interesting findings regarding sentiment of news headlines and their use in online news. These results bring relevant contributions to other researchers investigating characterization or prediction patterns of online news, to the design of online news systems and ultimately to society.
As a final contribution we have also made the dataset we built available in the hope that it can facilitate different research efforts in this field.

\section{Acknowledgments}
This work was funded by grants from CNPq, CAPES and FAPEMIG.

\balance

%
\bibliographystyle{aaai}
\small{\bibliography{arxiv-news-lightweight}} 

%
%

\end{document}